\documentclass[12pt]{article} 
\expandafter\let\csname equation*\endcsname\relax
\expandafter\let\csname endequation*\endcsname\relax
\usepackage{amsmath, euler}

\newcommand{\fl}{ \ }

\date{ \ }

\title{ Notes on the phase space formulation of the propagator of Hamiltonians with spatially-dependent kinetic energy}

\author{Yamen Hamdouni  \footnote{Email: hamdouniyamen@gmail.com} \\
\small Department of Physics, Faculty of Exact Sciences, \\ \small Mentouri University, Constantine,  Algeria 
}

\begin{document}
\maketitle

\begin{abstract}
These short notes present to the reader (students, in particular) a concise approach to the derivation of the propagator of Hamiltonians with 
position-dependent kinetic energy. The formalism is applied to the von Roos Hamiltonian with  arbitrary ordering ambiguity parameters, and 
a  simple scheme to convert the problem to a constant-mass motion is presented. The motion in curved spaces is treated along the same lines, where  a phase space  formulation 
is used to derive the propagator for arbitrary discretization choices.
\end{abstract}
\newpage
\section{Introduction}

The description of the dynamics of quantum systems is a central topic in modern physics which has attracted  a great deal of interest soon after 
the foundations of the quantum theory have been established. In the non-relativistic domain, the basic tool that enables for a such description resides 
in the Schr\"odinger equation \cite{quant}. 
The latter has been extended to the relativistic domain yielding thus the Klein-Gordan and the Dirac equations \cite{bjor}. Mathematically speaking,  the above equations
are second-order
partial differential
equations that may be solved exactly in some cases; this is in particular  true for many  potentials entering the Hamiltonian of
the Schr\"odinger equation.

Feynman came out with another, but equivalent, formalism known as the path integral formulation of the quantum mechanics \cite{feyn}. The latter concept has been successfully 
applied to many problems in physics, such as those in connection with  field theory and  condensed matter physics, to mention a few \cite{chaich,kl1,grosche,bastia}. The particular character of the path integral rests in the 
fact that there is no need to solve partial differential equations. Rather, one proceeds by evaluating the so-called propagator which makes it possible to link
the wave function
to its preceding history.  

Both the operator and the path integral methods apply well to systems for which the kinetic energy part of the Hamiltonian is independent of the position. However, as soon as
the dependence on the position is considered in the kinetic energy, the problem of the ordering and the discretization ambiguities immediately comes to play \cite{chaich,kl1}.  Actually, these ambiguities are  intimately related to each other since
the discretization ambiguity is a mere translation of the operator ordering problem inherent to the Hamiltonian, which 
 is a direct consequence of the non commuting character of the operators in quantum mechanics. Among the most studied systems displaying spatial dependence 
in the kinetic energy, we mention the motion of particles with 
position dependent effective masses, and the motion in curved spaces. In particular, motivated by several applications in atomic and condensed matter physics \cite{ap1,ap2,ap3,ap4,ap5,ap6,ap7,ap8,ap9,ap10,ap11,ap12}, 
a large number of works  has been devoted to finding  the solutions of
the Schr\"odinger equation for position-dependent effective mass; in this case,  one is usually  interested in finding analytical results for the wave functions 
and the energy spectrum corresponding to 
different forms of the mass, the potential and   the ordering ambiguity \cite{th1,th2,th3,th4,th5,th6,th7,th8,th9,th10,th11,th12,th13,th14}. The latter  is often encoded in the general form of the Hamiltonian initially proposed
by von Roos \cite{roos}.

On the other hand, some investigations have dealt with the application of the path integral formalism to the aforementioned problems \cite{th1,th2,th3,chet}. In the case of the von Roos Hamiltonian, one 
usually begins by adopting  a particular choice of 
the ordering, then proceeds to derive the discretized version of the path integral. However, so far, there is no general approach that takes into account
all the possible values of the ambiguity parameters in a concise and compact manner. The aim of this manuscript is to  derive explicit expressions of the 
propagator for arbitrary values of these  parameters, and to extend the investigation to the motion in curved space which turns out to be quite close to the
problem of spatially-dependent mass.

The plan of the discussion is as follows: In section~\ref{sec2}, we derive a compact formula for the action using a generalized definition of
the symbol of the Hamiltonian in phase space. Section~\ref{sec3} is devoted to the application of the obtained result
to the von Roos Hamiltonian, where we investigate different choices of the ordering parameters, and show how to reduce the propagator to that of a constant mass, 
generalizing thus the approach  of Alhaidari. In section~\ref{sec4} we deal with the derivation of the propagator in  curved spaces of arbitrary metrics.
We end the manuscript with a brief conclusion.

\section{Path integral formulation \label{sec2}}

Since our main concern is to formulate the propagator in terms of the phase space variables $\vec q$ and $\vec p$ corresponding to the particle in question, we   associate with every operator 
$ A$, a symbol $\mathcal A_\theta(\vec q, \vec p)$ that  depends on  $\vec q$ and $\vec p$ as follows:
\begin{equation}
 \langle \vec q_2 |A|\vec q_1\rangle =\frac{1}{(2\pi\hbar)^{3/2}}\int  \mathcal A_{\theta}\Bigl((1-\theta)\vec q_2+\theta \vec q_1,\vec p\Bigr)\langle
 \vec q_2-\vec q_1|\vec p\rangle d\vec p,   \label{trans1}
\end{equation}
where $\theta$ is a real parameter such that $0\leq \theta \leq 1$.  As we shall see bellow, it is this parameter that enables one to account for the different
discretization schemes  in the (lattice) formulation of the path integral. Notice that when $\theta=\frac{1}{2}$, we recover the Wigner transform of the operator $ A$ \cite{wigner}. Moreover,
 it is a matter of direct calculation to show that the inverse transformation corresponding to (\ref{trans1}) is given by
\begin{equation}
 \mathcal A_{\theta}(\vec q, \vec p)=(2\pi\hbar)^{3/2}\int\langle \vec q-\theta \vec q_1 |A|(1-\theta)\vec q_1+\vec q\rangle\langle \vec q_1|\vec p\rangle d\vec q_1 
  \label{trans2}.
\end{equation}

The evolution of a non-relativistic quantum particle whose Hamiltonian is $H$ and position is $\vec q$,  is described by its state vector $\psi$ which satisfies the Schr\"odinger equation:
\begin{equation}
 i\hbar \frac{\partial \psi}{\partial t}=H \psi.
\end{equation}
The solution of the latter equation can be formulated by means of the propagator $K(\vec q_f, t_f;\vec q_i,t_i)$ as 
\begin{equation}
 \psi(\vec q_f,t_f)=\int K(\vec q_f, t_f,\vec q_i,t_i) \psi(\vec q_i,t_i)d\vec q_i,
 \end{equation}
where $\psi(\vec q,t)=\langle \vec q,t|\psi\rangle$, and the kernel $K(\vec q_f, t_f;\vec q_i,t_i)$ is given in terms of the time-evolution operator of the system  by:

\begin{equation}
  K(\vec q_f, t_f;\vec q_i,t_i) =\langle \vec q_f,t_f|\vec q_i,t_i\rangle=\langle \vec q_f|U(t_f,t_i)|\vec q_i\rangle.
\end{equation}
From here one, we assume that $H$ does not depend explicitly on time. Then, by the Trotter formula one can write
\begin{eqnarray}
 \fl \langle \vec q_f|U(t_f,t_i)|\vec q_i \rangle&=&\lim_{N\to \infty}\langle q_f\bigl|\Bigl(1-\frac{i H }{\hbar N}(t_f-t_i)\Bigr)^N\bigl|q_i\rangle \nonumber \\
 &=&\lim_{N \to \infty}\int d \vec q_{N-1}\int d\vec q_{N-2}\cdots \int d\vec q_1 T_{\vec q_f \vec q_{N-1}} T_{\vec q_{N-1 }\vec q_{N-2}} \cdots T_{\vec q_1 \vec q_{0}},
\end{eqnarray}
with 
\begin{equation}
 T_{\vec q_k \vec q_{q-1}}=\langle \vec q_k\bigl|1-\frac{i H \epsilon }{\hbar}|\vec q_{k-1}\rangle. \label{prap}
\end{equation}
In the above we have set $\vec q_f=\vec q_N$, $q_i=\vec q_0$, $\epsilon=(t_f-t_i)/N$, and we have used the closure relation 
\begin{equation}
 \int |\vec q\rangle \langle \vec q| d\vec q=1. 
\end{equation}

By  using the transformation~(\ref{trans1}), together with the fact that
\begin{equation}
 \langle \vec q|\vec p\rangle=\frac{1}{(2\pi\hbar)^{3/2}}e^{\frac{i}{\hbar}\vec q \vec p},
\end{equation}
we can rewrite equation  (\ref{prap}) in the form
\begin{equation}
 T_{\vec q_k \vec q_{q-1}}=\frac{1}{(2\pi\hbar)^{3}}\int d\vec p_{k-1} e^{\frac{i}{\hbar}(\vec q_k-\vec q_{k-1}) \vec p_{k-1}}\Bigl(1-\frac{i \epsilon}{\hbar
 }\mathcal H_\theta\bigl((1-\theta)\vec q_k+\theta \vec q_{k-1}, \vec p_{k-1})\Bigr).
\end{equation}
Since $\epsilon$ is a very small quantity (in the end we shall take the limit $N\to\infty$) we can ascertain that 
\begin{equation}
 T_{\vec q_k \vec q_{q-1}}\approx \frac{1}{(2\pi\hbar)^{3}}\int d\vec p_{k-1} \exp\Bigl\{\frac{i\epsilon}{\hbar }\Bigl[\Bigl(\frac{\vec q_k-\vec q_{k-1}}{\epsilon}\Bigr)\vec p_{k-1}
-\mathcal H_\theta\bigl((1-\theta)\vec q_k+\theta \vec q_{k-1}, \vec p_{k-1})\bigr)\Bigr]\Bigr\}.
\end{equation}
It follows that the propagator  can be expressed as
\begin{eqnarray}
 \fl \langle \vec q_f|U(t_f,t_i)|\vec {q_i} \rangle &=&\lim_{N \to \infty}\iint d \vec q_{N-1} \frac{d\vec p_{N-1}}{(2\pi\hbar)^{3}}\iint d\vec q_{N-2}
 \frac{d \vec p_{N-2}}{(2\pi\hbar)^{3}}\cdots \iint d\vec q_1 \frac{d \vec p_{1}}{(2\pi\hbar)^{3}} \nonumber \\ \fl & \times &   \int \frac{d \vec p_{0}}{(2\pi\hbar)^{3}} 
 \exp\Biggl\{  \frac{i\epsilon}{\hbar }\sum\limits_{k=0}^{N-1}\Biggl[\Bigl(\frac{\vec q_{k+1}-\vec q_{k}}{\epsilon}\Bigr)\vec p_{k}
\nonumber \\&-&\mathcal H_\theta\bigl((1-\theta)\vec q_{k+1}+\theta \vec q_{k}, \vec p_{k})\bigr)\Biggr]\Biggr\}.\label{orig}
\end{eqnarray}
One can clearly notice that the discretization is encoded in the value of the parameter $\theta$: When $\theta=0$, we obtain a path integral with the postpoint 
discretization, whereas when $\theta=\frac{1}{2}$, we get a path integral with the midpoint discretization; the case $\theta=1$ yields a prepoint discretization.

 We can now put the propagator in a compact form as follows:
\begin{equation}
  K(\vec q_f, t_f;\vec q_i,t_i) =\int \mathcal D \vec q \int  \frac{\mathcal D \vec p}{(2\pi \hbar)^{3}} 
  \exp\Biggl\{\frac{i}{\hbar}\int\limits_{t_i}^{t_f} dt \bigl(\vec p \ \dot{\vec q}-\mathcal H_\theta(\vec q,\vec p)\bigr)\Biggr\}.
\end{equation}
The latter propagator describes the motion of a particle whose action in the phase space is given by 
\begin{equation}
 S=\int\limits_{t_i}^{t_f} dt \bigl(\vec p \ \dot{\vec q}-\mathcal H_\theta(\vec q,\vec p)\bigr).
\end{equation}
Hence it suffices to determine the explicit form of the phase space function $\mathcal H_\theta$ in order to find the expression of the propagator $K$. 
 In the following section, we shall apply  the above approach to a special form of the Hamiltonian, namely the von Roos Hamiltonian.
\section{Case of the von Roos Hamiltonian \label{sec3}}
\subsection{General form of the propagator}
The von Roos Hamiltonian was first introduced in order to account for the ordering  ambiguities that arise when one tries to quantize the motion of a spacially-dependent mass,
which we denote by $m(\vec q)$, under the effect of a  potential $V(|\vec q|)$. It has the form \cite{roos}
\begin{equation}
 H=\frac{1}{4}\bigl(m(\vec q)^\alpha  \vec{ p} \ m(\vec q)^\beta  \vec{ p} \ m(\vec q)^\gamma+m(\vec q)^\gamma  \vec{ p} \ m(\vec q)^\beta  \vec{ p} 
 \ m(\vec q)^\alpha\bigr) +V(|\vec q|),
\end{equation}
where  the mass has to be dealt with as an operator that does not commute with the momentum operator $\vec p$. The parameters $\alpha$, $\beta$ and $\gamma$ are real and
satisfy the obvious condition $\alpha +\beta+\gamma=-1$, which ensures  the correct form of the classical analog of $H$. 

By using the commutation relations satisfied by
the position and momentum operators, the Hamiltonian can be recast as
\begin{equation}
 H=\frac{1}{2m}{\vec p}^2+\frac{i\hbar}{2}\frac{\vec \nabla m}{m^2}\vec p-\frac{\hbar^2}{4m^3}\biggl[(\alpha+\gamma) m\Delta m-2(\alpha
 \gamma+\alpha+\gamma) \vec \nabla m.  \vec \nabla m\biggr]\label{hamil}.
\end{equation}
To construct the propagator, we first need to calculate the symbol $\mathcal H_\theta$ corresponding to~(\ref{hamil}). For this reason we write $H$ as
 the sum of three terms, namely,
 \begin{equation}
  H=H^1+H^2+H^3,
 \end{equation}
with
 \begin{eqnarray}
  H^1 &=&\frac{1}{2m}{\vec p}^2, \\ 
  H^2 &=& \frac{i\hbar}{2}\frac{\vec \nabla m}{m^2}\vec p, \\
  H^3 &=& -\frac{\hbar^2}{4m^3}\biggl[(\alpha+\gamma) m\Delta m-2(\alpha
 \gamma+\alpha+\gamma) \vec \nabla m.  \vec \nabla m\biggr]+V(|\vec q|).
 \end{eqnarray}
 Consequently it follows using equation~(\ref{trans2}) that
 \begin{eqnarray}
  \mathcal H^1_\theta(\vec q, \vec p)&=&\int d\vec \xi \ e^{\frac{i}{\hbar }\vec \xi \vec p}\langle \vec q-\theta \vec \xi|\frac{1}{2m(\vec q)}{\vec p}^2|
  \vec q+(1-\theta) \vec \xi\rangle \nonumber \\
  &=& \int d\vec \xi \ e^{\frac{i}{\hbar }\vec \xi \vec p} \frac{1}{2m(\vec q-\theta \vec \xi)}\langle \vec q-\theta \vec \xi|{\vec p}^2|
  \vec q+(1-\theta) \vec \xi\rangle.
 \end{eqnarray}
 By inserting the completeness relation for the momentum variable: 
 \begin{equation}
  \int d\vec {\tilde p} |\vec { {\tilde p}}\rangle\langle \vec {\tilde p} |=1,\label{close}
 \end{equation}
we obtain
\begin{equation}
  \mathcal{ H}^1_\theta(\vec q, \vec p)=\iint \frac{ d \vec {\tilde p}}{(2\pi\hbar)^3 } \frac{ d\vec \xi }{2m(\vec q-\theta \vec \xi)} \ 
  {\vec {\tilde p}}^2\ e^{\frac{i}{\hbar }\vec \xi (\vec p-\vec {\tilde p}) }.
  \end{equation}
  This equation can be further simplified on account of the properties of the Dirac delta function; hence, we may write
  \begin{equation}
  \mathcal H^1_\theta(\vec q, \vec p)=-\hbar^2 \int  \frac{ d\vec \xi }{2m(\vec q-\theta \vec \xi)} \ 
  e^{\frac{i}{\hbar } \vec \xi \vec p } \frac{\partial^2}{\partial {\vec{\xi}}^2} \ \delta(\vec \xi).
  \end{equation}
  Afterwards, integrating by parts, we find that
  \begin{equation}
  \mathcal H^1_\theta(\vec q, \vec p)=-\hbar^2 \lim_{\vec \xi \to 0}\frac{\partial^2} {\partial {\vec{\xi}}^2}  \Biggl(\frac{ 1 }{2m(\vec q-\theta \vec \xi)} \ 
  e^{\frac{i}{\hbar } \vec \xi \vec p }\Biggr).\label{massinv}
  \end{equation}
  It suffices now to carry on the partial derivation with respect to $\vec \xi$ to end up with 
  \begin{equation}
  \mathcal H^1_\theta(\vec q, \vec p)=\frac{{\vec p}^2}{2m(\vec q)}-2i\hbar \theta \frac{\vec \nabla m}{m^2} \vec p+
  \frac{\hbar^2 \theta^2}{m^3} \Bigl[m \Delta m -2 {\vec \nabla m}. {\vec \nabla m}\Bigr].
  \end{equation}
  Similarly, we can calculate $\mathcal H^2_\theta$ as follows:
\begin{eqnarray}
  \mathcal H^2_\theta(\vec q, \vec p)&=&\int d\vec \xi \ e^{\frac{i}{\hbar }\vec \xi \vec p}\langle \vec q-\theta \vec \xi|\frac{\vec \nabla m}{2m^2}{\vec p}|
  \vec q+(1-\theta) \vec \xi\rangle \nonumber \\
  &=& \int d\vec \xi \ e^{\frac{i}{\hbar }\vec \xi \vec p} \frac{\vec \nabla m|_{\vec q-\theta \vec \xi}}{2m(\vec q-\theta \vec \xi)^2}\langle \vec q-\theta \vec \xi|{\vec p}|
  \vec q+(1-\theta) \vec \xi\rangle 
 \end{eqnarray}
 Inserting the closure relation~(\ref{close}), we obtain
 \begin{eqnarray}
  \mathcal H^2_\theta(\vec q, \vec p)&=&\iint \frac{ d \vec {\tilde p}}{(2\pi\hbar)^3 } \frac{ d\vec \xi }{2m(\vec q-\theta \vec \xi)^2} \vec \nabla m|_{\vec q-\theta \vec \xi}\ 
  {\vec {\tilde p}}\ e^{\frac{i}{\hbar }\vec \xi (\vec p-\vec {\tilde p}) }\nonumber \\
  &=&-\frac{\hbar}{i}\int d\vec \xi e^{\frac{i}{\hbar }\vec \xi \vec p} \frac{ d\vec \xi }{2m(\vec q-\theta \vec \xi)^2} \vec \nabla m|_{\vec q-\theta \vec \xi}
   \frac{\partial}{\partial {\vec{\xi}}} \ \delta(\vec \xi).
   \end{eqnarray}
 Then,  integrating by parts yields:
   \begin{equation}
  \mathcal H^2_\theta(\vec q, \vec p)=\frac{\hbar}{i} \lim_{\vec \xi \to 0}\frac{\partial} {\partial {\vec{\xi}}}  
  \Biggl(\frac{ \vec \nabla m|_{\vec q-\theta \vec \xi} }{2m(\vec q-\theta \vec \xi)^2}  e^{\frac{i}{\hbar } \vec \xi \vec p } \Biggr).
    \end{equation}
   Carrying out partial derivation,  one finds that
    \begin{equation}
     \mathcal H^2_\theta(\vec q, \vec p)=\frac{\vec p \ \vec \nabla m}{2 m^2} +\frac{i\hbar \theta}{2m^3}\bigl[m\Delta m-2\vec \nabla m\vec \nabla m\bigr].
    \end{equation}
 Regarding the operator $H^3$, which depends only on the position, we simply have
\begin{equation}
 \mathcal H^3_\theta= -\frac{\hbar^2}{4m^3}\biggl[(\alpha+\gamma) m\Delta m-2(\alpha
 \gamma+\alpha+\gamma) \vec \nabla m.  \vec \nabla m\biggr]+V(|\vec q|), 
\end{equation}
where now $m$ is no longer an operator but a scalar that depends on the phase space variable $\vec q$.

Summing up all the contributions we finally obtain 
\begin{eqnarray}
 \mathcal H_\theta(\vec q, \vec p)&=&\frac{{\vec p}^2}{2m}+i\hbar \frac{\vec p \ \vec \nabla m}{2 m^2}(1-2\theta)+
  \frac{\hbar^2}{4m^3}\biggl\{\Bigl[2(\theta^2-\theta)-\alpha-\gamma\Bigr] m\Delta m \nonumber \\
  &-&2\Bigl[2(\theta^2-\theta)-\alpha
  \gamma-\alpha-\gamma \Bigr] \vec \nabla m.  \vec \nabla m\biggr\}+V(|\vec q|).
\end{eqnarray}
The latter symbol, together with the expression~(\ref{orig}) provide a unifying approach that takes into account both the ordering and the discretization ambiguities
of the path integral.

Let us briefly discuss  the general form of the propagator. Clearly, We can write the action  as
\begin{equation}
 S=S_{\rm Cl}+S_{Q}+i S_{\rm source},
\end{equation}
where $S_{\rm Cl}$ corresponds to the classical action, $S_{\rm Q}$ is the action of the quantum potential
\begin{equation}
 V_{\rm Q}=\frac{\hbar^2}{4m^3}\biggl\{\Bigl[2(\theta^2-\theta)-\alpha-\gamma\Bigr] m\Delta m 
  -2\Bigl[2(\theta^2-\theta)-\alpha
  \gamma-\alpha-\gamma \Bigr] \vec \nabla m.  \vec \nabla m\biggr\},
\end{equation}
and $S_{\rm source}$ is equivalent to the action of the source 
\begin{equation}
 \vec j=\hbar \frac{ \vec \nabla m}{2 m^2}(1-2\theta)
\end{equation}
that couples to the momentum ${\vec p}$. The above source vanishes for the midpoint discretization ($\theta=1/2$), and is nonzero for both the prepoint
and the postpoint discretization ($\theta=0,1$). Mathematically speaking, we may thus think of the motion as being equivalent to that of a particle under the effect of the potential $V+V_{\rm Q}$
which is disturbed in some sens by an imaginary  external source $\vec j$.
\subsection{Reduction to a constant mass problem}
\subsubsection{Propagator in one dimension}
From here on, in order to lighten the mathematical formulas,  we restrict ourselves to the motion in one dimension. In this case we can write the full propagator as
\begin{eqnarray}
 \fl K( q_f,t_f;q_i,t_i) &=&\lim_{N \to \infty}\iint d  q_{N-1} \frac{d p_{N-1}}{(2\pi\hbar)}\iint d q_{N-2}
  \frac{d  p_{N-2}}{(2\pi\hbar)}\cdots \iint d q_1 \frac{d p_{1}}{(2\pi\hbar)}\nonumber \\ & \times &   \int \frac{d  p_{0}}{(2\pi\hbar)} 
 \exp\Biggl\{  \frac{i\epsilon}{\hbar }\sum\limits_{k=0}^{N-1}\Biggl[\Bigl(\frac{ q_{k+1}- q_{k}}{\epsilon}\Bigr) p_{k}
-\frac{{ p_k}^2}{2m(\tilde q_k)}+i\hbar \frac{ p_k \  m'(\tilde q_k)}{2 m(\tilde q_k)^2}(1-2\theta)
  \nonumber \\ &- & \frac{\hbar^2}{4m(\tilde q_k)}\biggl\{\Bigl[2(\theta^2-\theta)-\alpha-\gamma\Bigr]  \frac{m(\tilde q_k)''}{m(q_k)}
 \nonumber \\ &-& 2\Bigl[2(\theta^2-\theta)-\alpha
 \gamma-\alpha-\gamma \Bigr]  \biggl(\frac{m(\tilde q_k)'}{m(\tilde q_k)}\biggr)^2\biggr\}-V(q)\Biggr] \Biggr\},
\end{eqnarray}
where $\tilde q_k=(1-\theta)q_{k-1}+\theta q_k$, and the prime denotes the derivation with respect to the position coordinate. We also confine the upcoming discussion to Weyl
ordering where $
 \theta=\frac{1}{2}$
which corresponds to the midpoint discretization of the path integral.  The propagator becomes
\begin{equation}
 K(q_f,t_f,q_i,t_i)=\iint \mathcal{D}q \frac{\mathcal{D}p}{{2\pi\hbar}}\exp\Bigl\{\frac{i}{\hbar}\int_{t_i}^{t_f}\bigl( p \dot q-\frac{p^2}{2m(q)}-
  V_{\rm eff}(q)\bigl)\Bigr\}\label{propfin}
\end{equation}
where the effective potential is given explicitly by
\begin{equation}
 V_{\rm eff}(q)=-\frac{\hbar^2}{8m( q)}\biggl\{\Bigl[1+2\alpha+2\gamma\Bigr] \frac{ m( q)''}{m(q)}
 - 2\Bigl[1+2\alpha
 \gamma+2\alpha+2\gamma \Bigr]  \biggl(\frac{m( q)'}{m( q)}\biggr)^2\biggr\}+V(q).
\end{equation}
Evidently, if we choose the von Roos  parameters such that 
\begin{equation}
 \alpha=\beta=-\frac{1}{2}, \qquad \gamma=0,
\end{equation}
or 
\begin{equation}
 \gamma=\beta=-\frac{1}{2}, \qquad \alpha=0,
\end{equation}
the quantum potential (that is the term proportional to $\hbar^2$) vanishes, and the  effective potential reduces to 
\begin{equation}
 V_{\rm eff}(q)=V(q).
\end{equation}
This shows that when the midpoint discretization is combined with the Li and Kuhn ordering \cite{li}, the resulting effective Hamiltonian is identical to the classical one.

However, 
if we choose $\alpha=\gamma=0$, which corresponds to the Ben Daniel and Duke ordering \cite{daniel}, we obtain
\begin{equation}
 V_{\rm eff}(q)=\frac{\hbar^2}{8}\frac{d^2}{dq^2}\frac{1}{m(q)}+V(q).
\end{equation}
In the case of Zhu and Kroemer ordering \cite{zhu}, that is when $\alpha=\gamma=-\frac{1}{2}$ and $\beta=0$, one finds
\begin{equation}
  V_{\rm eff}(q)=\frac{\hbar^2}{8m( q)}\biggl\{ \frac{ m( q)''}{m(q)}
 -  \biggl(\frac{m( q)'}{m( q)}\biggr)^2\biggr\}+V(q).
 \end{equation}
 \subsubsection{A simple scheme to convert the path integral to that of a constant mass} 
 
 The path integral derived above contains the spatially-dependent mass in the kinetic energy term. One usually uses a time transformation together with 
 regularizing functions \cite{chaich, klein}  to convert it to a path integral with constant mass.
 
 We propose another  simple  method to accomplish this task which generalizes  that  proposed by Alhaidari \cite{th7} for any ordering, and avoids the use of the Green's function.
 The essence of the proposed scheme may be summarized as follows. First consider the  representation of the Kernel in the
 basis formed by the eigenfunctions of the operator $H$, namely,
 \begin{equation}
   K(q_f,t_f,q_i,t_i)=\sum_n e^{-\frac{i}{\hbar}E_n(t_f-t_i)} \psi_n(q_f,t_f)\psi^*_n(q_i,t_i).
   \end{equation}
 Suppose now that:
 \begin{equation}
  \psi_(q,t)=\phi(q,t)F(q)^{1/4},
 \end{equation}
where $F(q)$ is a function of the position. Then we may write that
\begin{equation}
 [F(q_f)F(q_i)]^{-1/4}  K(q_f,t_f,q_i,t_i)=\widetilde  K(q_f,t_f,q_i,t_i)\label{extra}
\end{equation}
where 
\begin{equation}
 \widetilde  K(q_f,t_f,q_i,t_i)=\sum_n e^{-\frac{i}{\hbar}E_n(t_f-t_i)} \phi_n(q_f,t_f)\phi^*_n(q_i,t_i).
\end{equation}
Next let us consider the propagator in cartesian coordinate $x$:
\begin{equation}
 \bar K(x_f,t_f,x_i,t_i)=\iint \mathcal{D}x \frac{\mathcal{D}p}{{2\pi\hbar}}\exp\Bigl\{\frac{i}{\hbar}\int_{t_i}^{t_f}\bigl( p \dot x-\frac{p^2}{2}-\tilde V(x)\bigl)\Bigr\}
\end{equation}
and make the transformation
\begin{equation}
 x=F(q)
\end{equation}
Then it can be shown by expanding about the midpoint that the propagator can be written as~\cite{chaich}
\begin{eqnarray}
 \fl \bar K(q_f,t_i,q_i,t_i)&=&[F(q_f)F(q_i)]^{-1/4} \iint \mathcal{D}q \frac{\mathcal{D}p}{{2\pi\hbar}}\exp\Bigl\{\frac{i}{\hbar}\int_{t_i}^{t_f}\bigl
 ( p \dot q-\frac{p^2}{2F^{'}(q)^2}\nonumber \\ &-&\tilde V(F(q))-\frac{\hbar^2}{8}\frac{F''^2(q)}{F'^4(q)}\bigl)\Bigr\}\label{aftra} \label{proptr}.
\end{eqnarray}
Thus comparing equations~(\ref{proptr}) and (\ref{propfin}), it is sufficient on account of equation~(\ref{extra}), to choose
\begin{equation}
 F'(q)^2=m(q)
\end{equation}
which enables us to  map  the 
motion to that corresponding to a unit mass whose Hamiltonian is given by
\begin{equation}
  H_{\rm red}=\frac{ p^2}{2}+V_{\rm eff}(x)-\frac{\hbar^2}{32}\frac{m'(x)^2}{m(x)^3}=\frac{p^2}{2}+V_{\rm red}(x)
\end{equation}
where
\begin{equation}
 x=\int^q \sqrt{m(q')}dq',\label{pct}
\end{equation}
and
\begin{eqnarray}
 V_{\rm red}(x)&=&-\frac{\hbar^2}{8m( x)}\biggl\{\Bigl[1+2\alpha+2\gamma\Bigr] \frac{ m( x)''}{m(x)} \nonumber \\
 &-& 2\Bigl[\frac{7}{8}+2\alpha
 \gamma+2\alpha+2\gamma \Bigr]  \biggl(\frac{m( x)'}{m( x)}\biggr)^2\biggr\}+V(x)\label{vred}.
\end{eqnarray}
The above formulas can easily be adapted to the case where the reduced mass is arbitrary (i.e. not necessarily a unit mass, see the illustrative examples given bellow). 
Notice also that when $\alpha=\gamma=0$, that is for the Ben Daniel and Duke ordering,   we recover the result of Alhaidari.  

\subsubsection{Illustrative examples}

Let us illustrate the use of the above method by considering some particular forms of the mass. As a first example, we take
\begin{eqnarray}
 m(q)&=&m_0 /(a q)^{2},\nonumber \\
 V(q)&=&V_0,
\end{eqnarray}
where $m_0$, $a$ and $V_0$ are constants. Then from equation~(\ref{vred}), one finds  that for the Ben Daniel and Duke as well as for the Zhu and Kroemer orderings:
\begin{equation}
 V_{\rm red}=\frac{\hbar^2 a^2}{8m_0}+V_0.
\end{equation}
In the case of the Li and Kuhn ordering, we obtain
\begin{equation}
 V_{\rm red}=-\frac{\hbar^2 a^2}{8m_0}+V_0.
\end{equation}
All of the above orderings yield a constant potential, meaning that the wave function in the cartesian coordinate $x$ are merely 
of the form
\begin{equation}
 \psi(x)=A e^{-ik x}+B e^{ik x}
\end{equation}
where $A$ and $B$ are constants, and $k=\sqrt{2m_0(E-V_{\rm red})}/\hbar$,  with $E$ being the energy of the particle.
But by virtue of equation~(\ref{pct}):
\begin{equation}
 x=\frac{1}{a}\ln(a q),
\end{equation}
which leads to
\begin{equation}
 \psi(x)=A \exp\{-i\tfrac{k}{a} \ln(a q)\}+B \exp\{i \tfrac{k}{a}  \ln(a q)\}.
\end{equation}
A somewhat close form of this mass has been investigated by Schmidt in reference \cite{th10}  in the context of the revival of the wave function. There,  he also considered the form
\begin{equation}
 m(q)=x^\rho/\tau^2
\end{equation}
with $\rho\neq 2$. In this case, we obtain that for the Zhu and Kroemer ordering 
\begin{equation}
 V_{\rm red}(x)=-\frac{\hbar^2 \tau^2\rho(\rho+4)}{8(\rho+2)^2 x^2}.
\end{equation}
On the other hand for the Ben Daniel and Duke ordering, it turns out that
\begin{equation}
 V_{\rm red}(x)=\frac{\hbar^2 \tau^2\rho(3\rho+4)}{8(\rho+2)^2 x^2}.
\end{equation}
Similarly, we find that for the Li and Kuhn ordering,
\begin{equation}
 V_{\rm red}(x)=-\frac{\hbar^2 \tau^2}{8(\rho+2)^2 x^2}.
\end{equation}
All the above orderings yield a centrifugal  potential; however, in contrast to the Ben Daniel and Duke ordering which leads to a repulsive barrier, those corresponding
to Li and Kuhn, or Zhu and Kroemer produce an attractive barrier, leading thus to a completely different behavior of the motion of the particle.
It is worthwhile mentioning that the wave function in this case  is  given by  the Bessel function of the first kind, namely,
\begin{equation}
 \psi(x)=A \sqrt{x}  J_\nu(\sqrt{2E} x/2\hbar)
\end{equation}
where $E$ is the energy of the particle, and
\begin{equation}
 \nu=\begin{cases}
      \frac{1}{\rho+2} \qquad \qquad \qquad \text{ for Zhu-Kroemer}\\
       \frac{1}{2}\Bigl[1-\frac{1}{(\rho+2)^2 }\Bigr]^{1/2}      \quad \text{ for Li-Kuhn} \\
       \frac{\rho+1}{\rho+2}  \qquad \qquad \qquad \text{ for Ben Daniel-Duke} 
       
     \end{cases}
\end{equation}
To recover  the $q$-dependence,  we simply need to substitute 
\begin{equation}
 x=\frac{2 q^{1+\rho/2}}{\tau (\rho+2)}.
\end{equation}
in the expression of the wave function.

\section{Propagator in curved space \label{sec4}}
In this section we intend to extend the investigation to curved spaces. Recall that 
the Hamiltonian of a particle moving in a curved space with metric tensor $g_{\mu\nu}$ is given by \cite{bastia}
\begin{equation}
\hat H=\frac{1}{2} g^{-\frac{1}{4}}\hat p_\mu  g^{\mu\nu} g^{\frac{1}{2}}\hat p_\nu  g^{-\frac{1}{4}}+V( q)
\end{equation}
with $g={\rm det} g_{\mu\nu}$. This problem (initiated by DeWitt \cite{dewitt}) has been investigated by Mizrahi  \cite{miz}  in the case of the Weyl ordering. There the metric tensor has been dealt with as an ordinary 
operator in rectangular coordinates. Because of the curvature of the space, it would be preferable to follow another path.

To begin we emphasize  that  
to find the expression of the   propagator describing the motion of the particle, we  need to calculate the symbol $\mathcal H_\theta$  associated with the above 
Hamiltonian. The direct use of equation~(\ref{trans1}) is not possible in this case since the points $q-\theta q'$ and $q+(1-\theta)q'$ 
do not necessarily lie on a geodesic in the curved space of the particle. This problem does not come to play in the cartesian space since the geodesic is merely 
a straight line. Since our main aim is to deduce  the correct  transformation, it should be noted first of all
that the eigenvectors of the operator $\hat p_\mu$ 
may be calculated with the help of the expression $\hat p_\mu|p\rangle=\frac{\hbar}{i}(\partial_\mu+\frac{1}{2}\Gamma^\nu_{\nu\mu})|p\rangle$
 where $\Gamma^\nu_{\nu\mu}$ is the contracted affine connection.  But since $ \Gamma^\nu_{\nu\mu}=\frac{1}{2} \partial_\mu \ln g$,         
it follows that 
\begin{equation}
 \langle q|p\rangle=\frac{e^{\frac{i}{\hbar} qp}}{(2\pi\hbar)^{n/2}[g(q)]^{1/4}}\label{eigen}
\end{equation}
where $n$ is the dimension of the space. In this case the completeness relations in the position and the momentum spaces read:
\begin{eqnarray}
 1 &=& \int dq \sqrt{g(q)} |q\rangle\ \langle q|, \label{comp1}\\
 1 &=& \int dp  |p\rangle\ \langle p|. \label{comp2}
\end{eqnarray}

The symbol associated with the operator $\hat A$ has the general form 
\begin{equation}
 \mathcal A_\theta(q,p)=\int dq' Q_{\theta}(q,q') \langle q-\theta q'|\hat A|q+(1-\theta)q'\rangle e^{\frac{i}{\hbar} p q'}\label{gentr}
\end{equation}
where the weighting function $Q_\theta(q,q')$ is to be determined. The inclusion of the latter quantity is necessary to ensure covariance \cite{gnei}. 
Applying this to the operators $\hat q_\mu$ and $\hat p_\mu$ gives:
\begin{eqnarray}
 && Q_\theta(q,0)=\sqrt{g(q)},\nonumber \\
 && \frac{\hbar}{i}\lim_{q'\to0}\partial_\mu\Biggl(e^{\frac{i}{\hbar}pq'}\frac{Q_\theta(q,q')}{[g(q-\theta q')g(q+(1-\theta)q')]^{1/4}}\Biggl)=p_\mu.
 \end{eqnarray}
where the partial derivative is with respect to $q'$. This immediately yields
\begin{equation}
 Q_\theta(q,q')=[g(q-\theta q')g(q+(1-\theta)q')]^{1/4},
\end{equation}
which means that  equation~(\ref{gentr}) becomes
\begin{equation}
  \mathcal A_\theta(q,p)=\int dq' [g(q-\theta q')g(q+(1-\theta)q')]^{1/4} \langle q-\theta q'|A|q+(1-\theta)q'\rangle e^{\frac{i}{\hbar} p q'}.
\end{equation}
The latter formula enables us to calculate the symbol corresponding to the Hamiltonian of the particle as follows. First, one can see that
the last two terms of $\hat H$ depend only on the position, meaning that the corresponding 
symbol is simply obtained by replacing the position operators by the corresponding phase space variables. We only need to find the expression of the symbol of the term 
$\hat p_\mu  g^{\mu\nu}\hat p_\nu$, which is denoted in the subsequent discussion by  $\Bigl(\hat p_\mu  g^{\mu\nu}\hat p_\nu\Bigr)_\theta$.
Explicitly we have that:
\begin{eqnarray}
 \Bigl(\hat p_\mu  g^{\mu\nu}\hat p_\nu\Bigr)_\theta & =&\int dq'Q_\theta(q,q')\langle q-\theta q'|\hat p_\mu  g^{\mu\nu}\hat p_\nu|q+(1-\theta)q'\rangle 
 e^{\frac{i}{\hbar}pq'}\nonumber \\
 &=& \iiint\frac{ dq' dp' dp''}{(2\pi\hbar)^n} p'_\mu p''_\nu \langle p'| g^{\mu\nu}|p''\rangle e^{\frac{i}{\hbar}p'(q-\theta q')-p''(q+(1-\theta)q'))+pq'}.
\end{eqnarray}
Using  the completeness relation~(\ref{comp1}), we obtain
\begin{eqnarray}
 \Bigl(\hat p_\mu  g^{\mu\nu}\hat p_\nu\Bigr)_\theta & = & \iiiint dq'dp'dp'' dq''\frac{ \sqrt{g(q'')}}{(2\pi\hbar)^n}p'_\mu p''_\nu g^{\mu\nu}(q'')
 \langle p'|q''\rangle\nonumber \\
& \times & \langle q''|p''\rangle e^{\frac{i}{\hbar}p'(q-\theta q')-p''(q+(1-\theta)q'))+pq'}.
\end{eqnarray}
Afterwards, on account of  equation~(\ref{eigen}), we obtain after eliminating the variables $q''$, $ p'$ and $p''$
\begin{equation}
 \Bigl(\hat p_\mu  g^{\mu\nu}\hat p_\nu\Bigr)_\theta = -\hbar^2\int dq' e^{\frac{i}{\hbar}q'p} g^{\mu \nu}(q+(1-\theta)q')\partial_\mu\partial_\nu\delta(q').
\end{equation}
Whence,
\begin{equation}
 \Bigl(\hat p_\mu g^{\mu\nu}\hat p_\nu\Bigr)_\theta = -\hbar^2 \lim_{q'\to0}\partial_\mu\partial_\nu\Biggl( e^{\frac{i}{\hbar}q'p} g^{\mu \nu}(q+(1-\theta)q')\Biggr),
\end{equation}
where the derivation is with respect to $q'$. ( Notice the similarity that exists between the latter equation and equation~(\ref{massinv}).)
Now, performing the partial derivation, we find that
\begin{equation}
 \Bigl(\hat p_\mu  g^{\mu\nu}\hat p_\nu\Bigr)_\theta =g^{\mu \nu} p_\mu p_\nu+i\hbar\Bigl(\theta p_\mu \partial_\nu g^{\mu\nu}-(1-\theta) p_\nu \partial_\mu g^{\mu\nu}\Bigl)
 +\hbar^2\theta(1-\theta)\partial_\mu\partial_\nu g^{\mu\nu}.
\end{equation}
Let us focus our attention on the term that is proportional to $\hbar^2$. We need to calculate the derivative of the contravariant  metric tensor $g^{\mu\nu}$. To do so we make use
of the fact that $g^{\mu\nu}g_{\nu\sigma}=\delta^\mu_\sigma$, to find that
\begin{equation}
 \partial_\nu g^{\mu\sigma}=-\Bigl( g^{\mu\rho }\Gamma^\sigma_{\nu\rho}+ g^{\rho\sigma}\Gamma_{\rho\nu}^\mu\Bigr).
\end{equation}
Then,
\begin{eqnarray}
 \partial_\mu\partial_\nu g^{\mu\nu}&=&-[\partial_\nu g^{\mu\sigma} \Gamma^\nu_{\mu\sigma}+ g^{\mu\sigma}\Gamma^\nu_{\mu\sigma,\nu}+\partial_\nu g^{\nu\sigma}
 \Gamma^\mu_{\mu\sigma}+ g^{\sigma\nu}\Gamma^\mu_{\mu\sigma,\nu}]\nonumber \\
 &=&g^{\mu\rho}\Gamma^\sigma_{\nu\rho}\Gamma^\nu_{\mu\sigma}+g^{\rho\sigma}\Gamma^\nu_{\mu\sigma}\Gamma^\mu_{\nu\rho}-g^{\mu\sigma}\Gamma^\nu_{\mu\sigma,\nu}+
 g^{\rho\nu}\Gamma^\sigma_{\nu\rho}\Gamma_{\mu\sigma}^{\mu}\nonumber \\
 &+& g^{\rho\sigma}\Gamma^\mu_{\mu\sigma}\Gamma^\nu_{\nu\rho}-g^{\sigma\nu}\Gamma^\mu_{\mu\sigma,\nu}
\end{eqnarray}
where $\Gamma^\nu_{\mu\sigma,\rho}=\partial_\rho\Gamma^\nu_{\mu\sigma}$. Following the same method we can show that

\begin{eqnarray}
 \frac{1}{4}g^{1/4}\Delta g^{-1/4}&=&\frac{1}{4}g^{-1/4}\partial_\mu \Bigl(g^{1/4 }g^{\mu\nu}\Gamma^\rho_{\rho\nu}\Bigr)\nonumber \\
 &=& \frac{1}{4}g^{\mu\nu}\Bigl[\Gamma^\rho_{\rho\nu,\mu}-\frac{1}{2}\Gamma^\rho_{\rho\nu}\Gamma^\sigma_{\sigma\mu}-\Gamma^\rho_{\rho\sigma}\Gamma^\sigma_{\mu\nu}\Bigr].
\end{eqnarray}
Summing up all the terms, we end up with following expression of the symbol associated with the Hamiltonian $\hat H$:
\begin{eqnarray}
 \fl \mathcal H_\theta &=& \frac{1}{2}g^{\mu \nu} p_\mu p_\nu+\frac{i\hbar}{2}(1-2\theta) p_\nu
 \Bigl( g^{\mu\rho }\Gamma^\nu_{\mu\rho}+ g^{\rho\nu}\Gamma_{\mu\rho}^\mu\Bigr)+V(q)\nonumber \\
\fl  &+& \frac{\hbar^2}{2}g^{\mu\nu}\Biggl[\Bigl(\frac{1}{2}-\theta(1-\theta)\Bigr)\Bigl(\Gamma^\sigma_{\mu\sigma,\nu}-\Gamma^\sigma_{\mu\nu,\sigma}-
 \Gamma^\rho_{\rho\sigma}\Gamma^\sigma_{\mu\nu}+\Gamma^\rho_{\mu\sigma}\Gamma^\sigma_{\rho\nu}\Bigr)\nonumber \\
 \fl &+&\Bigl(\theta(1-\theta)-\frac{1}{4}\Bigr)\Gamma^\rho_{\rho\nu}\Gamma^\sigma_{\sigma\mu}+\Bigl(\frac{1}{2}-2\theta(1-\theta)\Bigr)\Gamma^\sigma_{\mu\nu,\sigma}
+\Bigl(3\theta(1-\theta)-\frac{1}{2}\Bigr)\Gamma^\rho_{\mu\sigma}\Gamma^\sigma_{\rho\nu}\Biggr].
\end{eqnarray}
Introducing the curvature 
\begin{equation}
 R=g^{\mu\nu}R_{\mu\nu}
\end{equation}
where $R_{\mu\nu}$ denotes the Ricci tensor:
\begin{equation}
 R_{\mu\nu}=\Gamma^\sigma_{\mu\sigma,\nu}-\Gamma^\sigma_{\mu\nu,\sigma}-
 \Gamma^\rho_{\rho\sigma}\Gamma^\sigma_{\mu\nu}+\Gamma^\rho_{\mu\sigma}\Gamma^\sigma_{\rho\nu},
\end{equation}
we finally find 
\begin{eqnarray}
\fl  \mathcal H_\theta(q,p)&=&\frac{1}{2}g^{\mu \nu} p_\mu p_\nu+\frac{i\hbar}{2}(1-2\theta) p_\nu
 \Bigl( g^{\mu\rho }\Gamma^\nu_{\mu\rho}+ g^{\rho\nu}\Gamma_{\mu\rho}^\mu\Bigr)+ \frac{\hbar^2}{2}\Bigl(\frac{1}{2}-\theta(1-\theta)\Bigr) R \nonumber \\ 
\fl  &+& \frac{\hbar^2}{2}g^{\mu\nu}\Biggl[
 \Bigl(\theta(1-\theta)-\frac{1}{4}\Bigr)\Gamma^\rho_{\rho\nu}\Gamma^\sigma_{\sigma\mu}  + \Bigl(\frac{1}{2}-2\theta(1-\theta)\Bigr)\Gamma^\sigma_{\mu\nu,\sigma} \nonumber \\
\fl &+&\Bigl(3\theta(1-\theta)-\frac{1}{2}\Bigr)\Gamma^\rho_{\mu\sigma}\Gamma^\sigma_{\rho\nu}\Biggr]+V(q).
\end{eqnarray}
The propagator can thus  be written for arbitrary values of the parameter $\theta$ as: 
\begin{eqnarray}
 \fl K(q',t';q,t)&=&[g(q)g(q')]^{-1/4}\lim_{N \to \infty}\iint d  q_{N-1} \frac{d p_{N-1}}{(2\pi\hbar)^{n}}\iint d q_{N-2}
  \frac{d  p_{N-2}}{(2\pi\hbar)^{n}}\cdots  \nonumber \\ \fl & \times &\iint d q_1 \frac{d p_{1}}{(2\pi\hbar)^{n}}   \int \frac{d  p_{0}}{(2\pi\hbar)^{n}} 
 \exp\Biggl\{  \frac{i\epsilon}{\hbar }\sum\limits_{k=0}^{N-1}\Biggl[\Bigl(\frac{ q_{k+1}- q_{k}}{\epsilon}\Bigr) p_{k} \nonumber \\
 &-&\mathcal H_\theta(\theta q_k+(1-\theta)q_{k+1},p_k)\Biggr]\Biggr\}.
\end{eqnarray}
Note finally that in the particular case of Weyl ordering, the symbol associated with the midpoint discretization reads
\begin{equation}
 \mathcal H_{\frac{1}{2}}=\frac{1}{2}g^{\mu \nu} p_\mu p_\nu+V(q)+\frac{\hbar^2}{8}\Bigl(R+g^{\mu\nu}\Gamma^\rho_{\mu\sigma}\Gamma^\sigma_{\rho\nu}\Bigr)
\end{equation}
as should be \cite{bastia, miz}.

\section{Concluding remarks}
 
We used a generalized definition of the phase space symbol to express the  action in a concise manner, and to  derive the propagator corresponding to  Hamiltonians with position-dependent  kinetic energy for different choices the associated inherent ambiguities. It turns out that  the lattice discretization ambiguity is encoded
in the value of  the parameter $\theta$. We have applied this formalism to the particular case of the von Roos Hamiltonian, with arbitrary values of the ordering
ambiguity parameters.
We find that in addition to the above quantities,  the lattice parameter $\theta$ also  determines the form of the quantum potential, providing thus
a unifying approach to the ordering and the discretization ambiguities. Moreover, we proposed a  simple
scheme that enables one to map the position-dependent mass problem to that of a constant mass, and we have demonstrated  how to use it by treating few simple examples.  
By extending the concept of the symbol to the case where the space is curved, we were able to express the propagator describing the motion in curved spaces.

 \end{document}